\documentclass[a4paper]{jpconf}
\usepackage[english]{babel}
\usepackage{amssymb,amsmath,amsthm,stmaryrd}
\usepackage{amsfonts}
\usepackage{euscript}
\allowdisplaybreaks[4]

\let\kappa=\varkappa
\let\phi=\varphi

\newcommand*{\eval}[1]{\left.#1\right|}
\newcommand{\pde}{\textsc{pde}}
\newcommand{\zcr}{\textsc{zcr}}
\newcommand{\ld}{\textsc{ld}}
\newcommand{\ES}[1]{\EuScript{#1}}

\DeclareMathOperator{\sym}{\mathbf{sym}}
\DeclareMathOperator{\cosym}{\mathbf{cosym}}
\DeclareMathOperator{\id}{id}

\newtheorem{proposition}{Proposition}
\theoremstyle{remark}
\newtheorem{remark}{Remark}

\begin{document}
\title{Higher symmetries of cotangent coverings for Lax-integrable
  multi-dimensional partial differential equations and Lagrangian
  deformations}

\author{H Baran$^1$, I S Krasil${}^\prime$shchik$^{1,2}$, O I Morozov$^3$ and
  P Voj{\v{c}}{\'{a}}k$^{1}$}

\address{$^1$ Mathematical Institute, Silesian University in Opava, Na
  Rybn\'{\i}\v{c}ku 1, 746 01 Opava, Czech Republic} \address{$^2$ Independent
  University of Moscow, B. Vlasevsky 11, 119002 Moscow, Russia} \address{$^3$
  Institute of Mathematics and Statistics, University of Troms\o, Troms\o \,
  9037, Norway}

\ead{josephkra@gmail.com}

\begin{abstract}
  We present examples of Lax-integrable multi-dimensional systems of partial
  differential equations with higher local symmetries. We also consider
  Lagrangian deformations of these equations and construct variational
  bivectors on them.
\end{abstract}

\section{Introduction}
Different approaches to integrability of partial differential equations (\pde
s), \cite{Mikhailov,Zakharov}, are based on their diverse but related
properties such as existence infinite hierarchies of (local or nonlocal)
symmetries and/or conservation laws, zero-curvature representations (\zcr),
bi-Hamiltonian structures, recursion operators, etc.

Much progress was achieved in the study of \pde s with two independent
variables. In particular, in a big number of examples it was shown that a
\pde\ integrable in the sense of presence of a \zcr\ with a non-removable
parameter (we call such equations \emph{Lax-integrable}) have infinite
hierarchies of higher local symmetries, see, e.g., \cite{Mikhailov}.  In the
multidimensional case the situation looks different. As far as we know, no
nontrivial examples of multi-dimensional equations with local higher
symmetries were found, cf.~\cite[\S 6]{Vinogradov1989}.

In this paper, we present five examples of multi-dimensional systems with
higher local symmetries. All these systems are defined as cotangent coverings,
\cite{KrasilshchikVerbovetsky2011,KrasilshchikVerbovetskyVitolo2012}, for
Lax-integrable \pde s with three or four independent variables, namely, for
the r-th dispersionsless Dym equation,
\cite{Blaszak,AlonsoShabat2004,Morozov2009,Ovsienko2010,Pavlov2003},
\begin{equation}
  u_{ty} = u_x\,u_{xy}-u_y\,u_{xx},\qquad
  \label{Ovsienko_eq}
  v_{ty} = u_x \, v_{xy} -u_y \, v_{xx} +2 \, (u_{xx} \, v_y -u_{xy} \, v_x),
\end{equation}
the r-th dispersionsless KP equation,
\cite{Blaszak,Dunajski2004,AlonsoShabat2004,OvsienkoRoger2008,Pavlov2003,Pavlov2006},
\begin{equation}
  u_{yy} = u_{tx}+u_y \, u_{xx}-u_x \, u_{xy},\qquad
  \label{Pavlov_eq}
  v_{yy} = v_{tx}+u_y \, v_{xx}-u_x \, v_{xy}+2 \, (u_{xy} \, v_x-u_{xx} \, v_y),
\end{equation}
the Veronese web equation,
\cite{BurovskyFerapontovTsarev2010,MarvanSergyeyev2012,Zakharevich},
\begin{eqnarray}\nonumber
  u_{xy} &=& \frac{\alpha \, u_x \, u_{ty}+(1-\alpha) \, u_y \, u_{tx}}{u_t},
  \\ \label{Veronese_web_eq}
  v_{xy} &=& \frac{\alpha \, u_x\, v_{ty}+\, (1-\alpha)\,u_y \, v_{tx}}{u_t}\\
  &&+ \frac{1}{u_t^2}\,
  ( (\alpha \, u_x \, u_{ty}+(1-\alpha) \, u_y \, u_{tx}) \, v_t
  \nonumber
  \\
  &&
  +( (1-2 \,\alpha) \,   u_t \, u_{ty}-(1- \alpha ) \, u_y \, u_{tt}) \, v_x
  -(\alpha \, u_x \, u_{tt}+(1-2 \,\alpha) \, u_t \, u_{tx}) \, v_y)
  \nonumber
  \\ \nonumber
  &&
  -\frac{v}{u_t^3} \,
  ((1-\alpha) \, u_y \, u_{tt} \, u_{tx}
  -u_t \, u_{tx} \, u_{ty}+\alpha \, u_x \, u_{tt} \, u_{ty})
\end{eqnarray}
with $\alpha \neq0, 1$, the universal hierarchy equation,
\cite{AlonsoShabat2004,Pavlov2003},
\begin{equation}
  u_{yy}  =  u_y \, u_{tx}-u_x \, u_{ty},\qquad
  \label{universal_hierarchy_eq}
  v_{yy}  =  u_y \, v_{tx} -u_x \, v_{ty}+ 2 \, (u_{ty} \, v_x-u_{tx} \, v_y),
\end{equation}
and for the 4-dimensional analogue of~\eqref{Ovsienko_eq},~\cite{Morozov2013},
\begin{equation}
  u_{ty}  =  u_z\,u_{xy}-u_y\,u_{xz},\qquad
  \label{rdDym4D_eq}
  v_{ty}  =  u_z \, v_{xy} -u_y \, v_{xz} +2 \, (u_{xz} \, v_y -u_{xy} \, v_z),
\end{equation}
introduced in \cite{AlonsoShabat2004}.  We show that these systems have local
symmetries of the third order.  Equations~\eqref{Ovsienko_eq},
\eqref{Pavlov_eq} and~\eqref{Veronese_web_eq} belong to the families of \pde s
\begin{gather}
  u_{ty} = u_x\,u_{xy}+ \kappa u_y\,u_{xx},
  \label{rdDym_eq}\\
  u_{yy} = u_{tx}+\left(\case{1}{2}\,(\kappa+1)\,u_x^2+u_y\right) \, u_{xx}+
  \kappa \,u_x \, u_{xy},
  \label{rmdKP_eq}\\
  u_{xy} = \frac{\alpha \, u_x \, u_{ty}+\beta \, u_y \, u_{tx}}{u_t}
  \label{generalized_Veronese_web_eq}
\end{gather}
and are featured as the only representatives of their families whose \zcr s
have a non-removable parameter. We make an observation that all the symmetries
of the cotangent coverings of Equations~\eqref{rdDym_eq} with $\kappa \not =
-1$, \eqref{rmdKP_eq} with $\kappa \not = -1$,
and~\eqref{generalized_Veronese_web_eq} with $\beta \not = 1 - \alpha$ up to
order 5 are the point symmetries. This leads to a conjecture that the
cotangent coverings of Equations~\eqref{rdDym_eq}, \eqref{rmdKP_eq},
\eqref{generalized_Veronese_web_eq} have local higher symmetries whenever
these equations are Lax-integrable.  If this claim turns out to be correct, it
will be a valuable refinement of the observations of \cite[\S
6]{Vinogradov1989}.

In the last section, we consider \emph{Lagrangian deformations} of cotangent
coverings to some of the equations considered above and construct variational
bivectors, see~\cite{KrasilshchikVerbovetsky2011}, on these deformations.

\section{Basics}\label{sec:basics}
We briefly recall here the basic definitions from the geometry of \pde s,
including the definition of the cotangent covering,
see~\cite{KrasilshchikVerbovetsky2011,KrasilshchikVerbovetskyVitolo2012}, and
its Lagrangian deformations.  Let~$\pi \colon \mathbb{R}^n \times \mathbb{R}^m
\to \mathbb{R}^n$ be the trivial bundle with the coordinates~$(x^1,\dots,x^n)$
in~$\mathbb{R}^n$ and~$(u^1,\dots,u^m)$ in~$\mathbb{R}^m$ and~$J^\infty(\pi)$
be the space of its infinite jets. Local coordinates on $J^\infty(\pi)$ are
$(x^i,u^\alpha,u^\alpha_I)$, where $I=(i_1, \dots, i_k)$ is a multi-index, and
for every local section $f$ of $\pi$ and the corresponding infinite jet
$j_\infty(f)\colon \mathbb{R}^n \to J^\infty(\pi)$ the coordinate~$u^\alpha_I$
is defined by $\eval{u^\alpha_I}_{j_\infty(f)} =\displaystyle{\frac{\partial
    ^{\#I} f^\alpha}{\partial x^I}} =\displaystyle{\frac{\partial
    ^{i_1+\dots+i_n} f^\alpha}{(\partial x^1)^{i_1}\dots (\partial
    x^n)^{i_n}}}$.  We put $u^\alpha = u^\alpha_{(0,\dots,0)}$.

The vector fields
\begin{equation*}
D_{x^k} = \frac{\partial}{\partial x^k} + \sum \limits_{\# I \ge 0} \sum
\limits_{\alpha = 1}^m u^\alpha_{I+1_{k}}\,\frac{\partial}{\partial
  u^\alpha_I}, \qquad k =1,\dots,n,
\end{equation*}
$(i_1,\dots, i_k,\dots, i_n)+1_k = (i_1,\dots, i_k+1,\dots, i_n)$, are called
\emph{total derivatives}.  They commute everywhere on $J^\infty(\pi)$.

The \emph{evolutionary derivation} associated to an arbitrary smooth function
$\varphi \colon J^\infty(\pi) \to \mathbb{R}^m $ is the vector field
\begin{equation}
  \mathbf{E}_{\varphi} = \sum \limits_{\# I \ge 0} \sum \limits_{\alpha = 1}^m
  D_I(\varphi^\alpha)\,\frac{\partial}{\partial u^\alpha_I}
  \label{evolution_differentiation}
\end{equation}
with $D_I=D_{(i_1,\dots\,i_n)} =D^{i_1}_{x^1} \circ \dots \circ
D^{i_n}_{x^n}$.

A system of \pde s $F_a(x^i,u^\alpha_I) = 0$, $\# I \le s$, $a=1,\dots, r$, of
order $s \ge 1$ with $r \ge 1$ defines the submanifold $\ES{E} =
\{(x^i,u^\alpha_I) \in J^\infty(\pi) \,\,\vert\,\, D_K(F_a(x^i,u^\alpha_I)) =
0, \,\, \# K \ge 0\}$ in $J^\infty(\pi)$.

A function $\varphi \colon \ES{E} \to \mathbb{R}^m$ is called a \emph{
  \textup{(}generator of an infinitesimal\textup{)} symmetry} of $\ES{E}$ when
$\mathbf{E}_{\varphi}(F) = 0$ on $\ES{E}$. The symmetry $\varphi$ is a
solution to the \emph{defining system}
\begin{equation}
  \ell_{\ES{E}}(\varphi) = 0,
  \label{defining_eqns}
\end{equation}
where $\ell_{\ES{E}} = \eval{\ell_F}_{\ES{E}}$ with the matrix
differential operator
\begin{equation*}
\ell_F = \left(\sum \limits_{\# I \ge 0}\frac{\partial F_a}{\partial
    u^\alpha_I}\,D_I\right).
\end{equation*}
Solutions to~\eqref{defining_eqns} constitute
the Lie
algebra with respect
to the \emph{Jacobi bracket} $\{\varphi, \psi\} =
\mathbf{E}_{\varphi}(\psi)-\mathbf{E}_{\psi}(\varphi)$ denoted
by~$\sym(\ES{E})$.  The subalgebra of \emph{contact symmetries} of $\ES{E}$ is
$\sym_0(\ES{E}) =\sym(\ES{E}) \cap C^\infty(J^1(\pi),\mathbb{R}^m)$.
Symmetries from $\sym(\ES{E}) \setminus C^\infty(J^1(\pi),\mathbb{R}^m)$ are
said to be \emph{higher} ones.

Dually, a \emph{cosymmetry} is a function~$\psi\colon \ES{E} \to \mathbb{R}^r$
that satisfies the equation
\begin{equation*}
  \ell_{\ES{E}}^*(\psi) = 0,
\end{equation*}
where~$\ell_{\ES{E}}^*$ denotes the formally adjoint operator
\begin{equation*}
  {\ell}_F^{*} = \left( \sum \limits_{\# I \ge 0} (-1)^{\# I} {D}_I
  \circ \frac{\partial F_a}{\partial u^\alpha_I} \right)^{T}
\end{equation*}
restricted to~$\ES{E}$.

A differential operator~$\Delta\colon C^\infty(\ES{E},\mathbb{R}^r)\to
C^\infty(\ES{E},\mathbb{R}^m)$ is called a \emph{variational bivector} if it
takes symmetries of the equation~$\ES{E}$ to its cosymmetries and
satisfies~$(\ell_{\ES{E}}\circ\Delta)^*=\ell_{\ES{E}}\circ\Delta$. Any such an
operator defines a skew-symmetric
bracket~$\{\omega_1,\omega_2\}_\Delta=L_{\Delta(\delta\omega_1)}(\omega_2)$ on
the space of conservation laws, where~$\delta=\delta^{0,n-1}$ is the
differential in the term~$E_1$ of the Vinogradov $\ES{C}$-spectral sequence
(see~\cite{Vinogradov:SSLFCLLTNT}) and~$L$ denotes the Lie derivative. If the
variational Schouten bracket~$\llbracket\Delta,\Delta\rrbracket$ vanishes
then~$\Delta$ is a Hamiltonian operator,
see~\cite{KrasilshchikVerbovetsky2011}.

Denote $\mathcal{W} = \mathbb{R}^l$, $l\leq\infty$, with coordinates
$w^0,\dots,w^s,\dots$. Locally, a \emph{differential covering} over $\ES{E}$ is
a trivial bundle $\tau \colon\tilde{\ES{E}}= \ES{E}\times \mathcal{W} \to
\ES{E}$ equipped with the \emph{extended total derivatives}
\begin{equation}
  \tilde{D}_{x^k} = D_{x^k} + \sum \limits_{s =0}^{l}
  T^s_k(x^i,u^\alpha_I,w^j)\,\frac{\partial }{\partial w^s}
  \label{extended_derivatives}
\end{equation}
such that $[\tilde{D}_{x^i}, \tilde{D}_{x^j}]=0$. This yields the
system of  the \emph{covering equations}
\begin{equation}
  w^s_{x^k} = T^s_k(x^i,u^\alpha_I,w^j).
  \label{WE_prolongation_eqns}
\end{equation}
This over-determined system of \pde s is compatible whenever $(x^i,u^\alpha_I)
\in \ES{E}$.

The \emph{cotangent covering} is defined as follows. Let, as
before,~$\ES{E}\subset J^\infty(\pi)$ be given by the
relations~$F_1=\dots=F_r=0$, where~$F_a=F(x^i,u_I^j)$. Consider the
bundle~$\pi'\colon \mathbb{R}^r\times\mathbb{R}^m\times\mathbb{R}^n\to
\mathbb{R}^n$ with coordinates~$v^\alpha$
in~$\mathbb{R}^r$. Then~$\ES{T}^*\ES{E}\subset J^\infty(\pi')$ is the equation
obtained from~$\ES{E}$ by adding the
relations~$\ell_{\ES{E}}^*(v^\alpha)=0$. The natural
projection~$\tau^*\colon\ES{T}^*\ES{E}\to\ES{E}$ is an infinite-dimensional
covering. Note that~$\ES{T}^*\ES{E}$ is always an Euler-Lagrange equation with
the Lagrangian density~$\ES{L}=
(F_1v^1+\dots+F_rv^r)\,dx^1\wedge\dots\wedge\,dx^n$. Consequently, the spaces
of symmetries~$\sym(\ES{E})$ and cosymmetries~$\cosym(\ES{E})$ coincide
for~$\ES{T}^*\ES{E}$.

\section{Higher symmetries of cotangent coverings}
\label{sec:high-symm-cotang}
In this section we present the results of computation of the third order
symmetries for Systems~\eqref{Ovsienko_eq}--\eqref{rdDym4D_eq}.  The
computations were held in the \textsc{Jets}
software,~\cite{Jet,Marvan2009}. We use the notation~$u^1= u$, $u^2=v$,
$x^1=t$, $x^2=x$, $x^3 = y$, and $x^4 = z$ in the formulas of the previous
section.

\begin{proposition}
  The cotangent coverings of the Lax-integrable cases of the r-th
  dispersionsless Dym equation\textup{,} r-th modified dispersionsless KP
  equation\textup{,} the Veronese web equation\textup{,} the universal hierarchy
  equation\textup{,} and the $4$-dimensional r-th dispersionsless Dym equation
  have the following symmetries up to the third order\textup{:}

  \textup{1)} System~\eqref{Ovsienko_eq}\textup{:}
  \begin{eqnarray*}
    \phi_0(A_0) &=& \left(A_0 \, u_t  + A_{0,t} \, (x\,u_x -u)+\case{1}{2} \,
      A_{0,tt} \, x^2,
      A_0 \, v_t+ A_{0,t} \, (x \, v_x+2 \,v)\right),
    \\
    \phi_1(A_1) &=& \left(A_1 \, u_x+A_{1,t} \, x, A_1 \, v_x\right),
    \\
    \phi_2(A_2) &=& \left(A_2, 0\right),
    \\
    \phi_3(A_3) &=& \left(0, A_3 \,( 2 \, u_t - 3 \,u_x^2) + A_{3,t} \,(2\, x\,
      u_x + u)
      -\case{1}{2}  \, A_{3,tt}\, x^2 \right),
    \\
    \phi_4(A_4) &=& \left(0, 2 \, A_4 \, u_x - A_{4,t} \, x\right),
    \\
    \phi_5(A_5) &=& \left(0, A_5\right),
    \\
    \phi_6(B_0) &=& \left(B_0 \, u_y, B_0 \, v_y\right),
    \\
    \phi_7(B_1) &=& \left(0, B_1\,u_y^{-2}\right),
    \\
    \phi_8 &=& \left(x \, u_x-2 \, u, x \, v_x\right),
    \\
    \phi_9 &=& \left(0, v\right),
    \\
    \phi_{10} &=& \left(0,
      u_{ttt}
      -3 \, u_x \, (u_{ttx}-u_x\, u_{txx}-u_{tx} \, u_{xx})
      -u_x^3 \, u_{xxx}
      -\case{3}{2} \, (u_{tx}^2+u_x^2 \, u_{xx}^2)
    \right),
    \\
    \phi_{11} &=& \left(0, u_{ttx}-2 \, u_x \, u_{txx}+u_x^2 \, u_{xxx}
      -u_{xx}\,(u_{tx} -u_x \, u_{xx})\right),
    \\
    \phi_{12} &=& \left(0, u_{txx}-u_x \, u_{xxx}-\case{1}{2} \, u_{xx}^2\right),
    \\
    \phi_{13} &=& \left(0, u_{xxx}\right),
    \\
    \phi_{14} &=& \left(0, (2 \,u_y \,u_{xxy}-u_{xy}^2)\,u_y^{-2}\right),
    \\
    \phi_{15} &=& \left(0, (u_y \, u_{xyy}-u_{xy} \, u_{yy})\,u_y^{-3}\right),
    \\
    \phi_{16} &=& \left(0, (2 \,u_y \, u_{yyy}-3 \, u_{yy}^2)\,u_y^{-4}\right),
    \end{eqnarray*}
    where $A_i=A_i(t)$\textup{,} $B_i=B_i(y)$ are arbitrary functions of their
    arguments\textup{;}

    \textup{2)} System~\eqref{Pavlov_eq}\textup{:}
    \begin{eqnarray*}
      \phi_0(A_0) &=&
      \left(A_0 \, u_t+(x \, u_x+y \, u_y-u)
        \,A_{0,t}+(\case{1}{2} \, u_x \, y^2-x \, y) \, A_{0,tt} 
      -\case{1}{6} \, y^3 \, A_{0,ttt},
        \right.
        \\
        &&
        \qquad\qquad
        \left.
         A_0 \, v_t+(x\,v_x  +y \,v_y +2 \, v) \, A_{0,t}
        +\case{1}{2}\, y^2  \, v_x \, A_{0,tt}
       \right),
      \\
      \phi_1(A_1) &=& \left(
        A_1 \,u_y + A_{1,t} \,(y\,u_x  -x)-\case{1}{2} \, A_{1,tt} \, y^2,
        A_1 \, v_y + A_{1,t}\, y \, v_x\right),
      \\
      \phi_2(A_2) &=& \left(A_2 \,u_x \, -A_{2,t} \, y, A_2 \,v_x\right),
      \\
      \phi_3(A_3) &=& \left(A_3, 0\right),
      \\
      \phi_4(A_4) &=& \left(0,
        2 A_4 \, (u_t +2 \, u_x^3+3  \, u_x \, u_y)
        + A_{4,t} \, (3 \, y \, u_x^2 + 2 \, x \, u_x
        +2 \, y \, u_y
        + u)
      \right.
      \\
      &&
      \qquad\qquad
      \left.
        + A_{4,tt} \,y\,(y \, u_x + x)
        +\case{1}{6} \, A_{4,ttt} \, y^3
      \right),
      \\
      \phi_5(A_5) &=& \left(0,
        A_5 \,( 2 \, u_y+3 \,  u_x^2)
        +A_{5,t} \, (2 \, y \, u_x+x)
        +\case{1}{2}\, A_{5,tt} \, y^2\right),
      \\
      \phi_6(A_6) &=& \left(0, 2  \, A_6 \, u_x+ A_{6,t} \, y\right),
      \\
      \phi_7(A_7) &=& \left(0, A_7\right),
      \\
      \phi_8 &=& \left(
        2\, x \, u_x +y \, u_y -3 \, u,
        2\, x \, v_x
        +y\, v_y
        + 4 \, v
        +2 \, t\, u_t
        + u_x \,(2 \, x +3 \, y \, u_x+4 \, t \, u_x^2)
      \right.
      \\
      &&
      \left.
        \qquad\qquad\qquad
        +2\, u_y \,(3 \, t \, u_x + y)
        +u
      \right),
      \\
      \phi_9 &=&\left(y \, u_x+2 \, t \, u_y,
        y \, v_x+2 \, t \, v_y\right),
      \\
      \phi_{10} &=& \left(0, v\right),
      \\
      \phi_{11} &=& \left(0, u_{xxx}\right),
      \\
      \phi_{12} &=& \left(0, u_{xxy}+u_x \, u_{xxx} +\case{1}{2} \,
        u_{xx}^2\right), 
      \\
      \phi_{13} &=& \left(0, u_{txx}
        +u_{xxx} \, (u_x^2+u_y)
        +u_x \, u_{xxy}
        +u_{xx} \, (u_{xy}+u_x\, u_{xx})
      \right),
      \\
      \phi_{14} &=& \left(0,
        2 \, u_{txy}
        +4 \, u_x \, u_{txx}
        +2 \, u_x \, u_{xxx} \, (u_x^2+2 \, u_y)
        +2 \, u_{xxy} \, (u_x^2+u_y)
        + u_{xx}^2 \,(3 \, u_x^2+2 \, u_y)
      \right.
      \\
      &&
      \qquad
      \left.
        +u_{xy}^2
        +2 \, u_{xx} \, (u_{tx} +2 \, u_x \, u_{xy})
      \right),
      \\
      \phi_{15} &=& \left(0, u_{ttx}
        + u_{txx}\,(3 \, u_x^2 + 2 \, u_y)
        +2 \, u_x \, u_{txy}
        +u_{xxx} \,(u_x^4+3 \, u_y \, u_x^2+u_y^2)
        +u_x \, u_{xy}^2
      \right.
      \\
      &&
      \qquad
      \left.
        +u_x \, u_{xxy} \, (u_x^2+2 \, u_y)
        +u_{tx} \,(u_{xy}+3 \, u_x \, u_{xx})
        +u_x \,u_{xx}^2\,(2 \, u_x^2 +3  \, u_y)
      \right.
      \\
      &&
      \qquad
      \left.
        +u_{xx} \, (u_{ty}+   u_{xy}\, (3 \, u_x^2+2 \, u_y))
      \right),
      \\
      \phi_{16} &=& \left(0,
        2 \, u_{tty}
        +6 \, u_x\, u_{ttx}
        +4 \,u_x \,u_{txx} \,(2\,u_x^2+ 3 \,u_y)
        +2 \,u_{txy} \,(2 \,u_y+ 3\,u_x^2)
      \right.
      \\
      &&
      \qquad
      \left.
        +2 \,u_x \,u_{xxx} \,(u_x^4+4\,u_x^2 \,u_y + 3\,u_y^2)
        +2 \,u_{xxy} \,(u_x^4+ 3\,u_x^2 \,u_y+u_y^2)
        +3 \, u_{tx}^2
      \right.
      \\
      &&
      \qquad
      \left.
        +6 \, u_{tx} \, (u_x \, u_{xy}+ u_{xx} \, (u_y+2 \,u_x^2))
        + u_{xx}^2 \,(5 \,u_x^4 +12 \,u_x^2 \,u_y + 3\,u_y^2)
      \right.
      \\
      &&
      \qquad
      \left.
        + u_{xy}^2 \,(3 \,  u_x^2 +2  \, u_y)
        +2 \, u_{ty} \, (u_{xy}+3 \, u_x \,  u_{xx})
        + 4 \,u_x \,u_{xx} \,u_{xy} \,(2 \,u_x^2 + 3 \,u_y)
      \right),
      \\
      \phi_{17} &=&\left(0,
        u_{ttt}
        +3 \, (2 \, u_x^2+u_y) \, u_{ttx}
        +3 \, u_x \, u_{tty}
        +(5 \, u_x^4+12 \, u_x^2 \, u_y+3 \, u_y^2) \, u_{txx}
      \right.
      \\
      &&
      \qquad
      \left.
        +(5 \, u_y \, u_x^4+6 \, u_y^2 \, u_x^2+u_x^6+u_y^3) \, u_{xxx}
        +u_x \, (u_y+u_x^2) \, (3 \, u_y+u_x^2) \, u_{xxy}
      \right.
      \\
      &&
      \qquad
      \left.
        +12 \,u_x   \, u_y \, u_{tx} \, u_{xx}
        +10 \, u_x^3 \, u_{tx}  \, u_{xx}
        +6\, u_x \, u_{tx}^2
        +2 \, u_x \, (2 \, u_x^2+3 \, u_y) \, u_{txy}
      \right.
      \\
      &&
      \qquad
      \left.
        +5 \, u_x^4\, u_{xx}  \, u_{xy}
        +12 \, u_x^2 \, u_y \, u_{xx} \, u_{xy}
        +3 \, u_y \, u_{tx}  \, u_{xy}
        +6\, u_x^2  \, u_{tx} \, u_{xy}
        +3\, u_x \, u_y \, u_{xy}^2
      \right.
      \\
      &&
      \qquad
      \left.
        +3\, u_y^2  \, u_{xx} \, u_{xy}
        +3\, u_x^5 \, u_{xx}^2
        +6\, u_x\, u_y^2  \, u_{xx}^2
        +10\, u_y\, u_x^3  \, u_{xx}^2
      \right.
      \\
      &&
      \qquad
      \left.
        +3 \, (u_{tx}+ 2 \, u_{xx} \, u_x^2+u_y \, u_{xx}+u_x \, u_{xy}) \, u_{ty}
        +2 \, u_x^3 \, u_{xy}^2\right),
    \end{eqnarray*}
    where $A_i=A_i(t)$ are arbitrary functions of $t$\textup{;}

    \textup{3)} System~\eqref{Veronese_web_eq}\textup{:}
    \begin{eqnarray*}
      \phi_{1}(P_1) &=& \left(P_1 \, u_t , P_1 \, v_t + P_{1,t} \,v \right),
      \\
      \phi_{2}(Q_1) &=& \left(Q_1 \, u_x, Q_1 \, v_x\right),
      \\
      \phi_{3}(R_1) &=& \left(R_1 \, u_y, R_1 \, v_y\right),
      \\
      \phi_{4}(S_1) &=& \left(S_1, -S_{1,u} \, v\right),
      \\
      \phi_{5} &=& \left(u, 0\right),
      \\
      \phi_{6}(P_2) &=& \left(0, P_2\,u_t^{-1}\right),
      \\
      \phi_{7}(Q_2) &=& \left(0, Q_2 \, u_t\,u_x^{-2}\right),
      \\
      \phi_{8}(R_2) &=& \left(0, R_2 \, u_t\,u_y^{-2}\right),
      \\
      \phi_{9}(S_2) &=& \left(0, S_2 \, u_t\right),
      \\
      \phi_{10} &=& \left(0,
        (2\, u_t \, u_x \, u_{ttx}- 2\, u_x\, u_{tt} \, u_{tx}-u_t \,
        u_{tx}^2) \,u_t^{-2}\,u_x^{-2}\right),
      \\
      \phi_{11} &=& \left(0,
        (2\, u_t \, u_y \, u_{tty}- 2\, u_y \, u_{tt} \, u_{ty}-u_t \,
        u_{ty}^2) \,u_t^{-2}\,u_y^{-2}\right),
      \\
      \phi_{12} &=& \left(0, (2 \,u_t \,u_x \, u_{txx}-u_x \, u_{tx}^2-
        2\,u_t\, u_{tx}\, u_{xx}) \,u_t^{-1} \,u_x^{-3}\right),
      \\
      \phi_{13} &=& \left(0, (2\,u_t \,u_y \, u_{tyy}-u_y \, u_{ty}^2-2\,u_t\,
        u_{ty}\, u_{yy})\,u_t^{-1}\,u_y^{-3}\right),
      \\
      \phi_{14} &=& \left(0, (2\, u_t \, u_{ttt}-3 \,
        u_{tt}^2)\,u_t^{-3}\right), 
      \\
      \phi_{15} &=& \left(0, u_t \, (2\, u_x \, u_{xxx}-3 \,
        u_{xx}^2)\,u_x^{-4}\right),
      \\
      \phi_{16} &=& \left(0, u_t \, (2\, u_y \, u_{yyy}-3 \,
        u_{yy}^2)\,u_y^{-4}\right),
    \end{eqnarray*}
    where $P_i=P_i(t)$\textup{,} $Q_i=Q_i(x)$\textup{,} $R_i=R_i(y)$\textup{,}
    $S_i=S_i(u)$ are arbitrary functions of their arguments\textup{;}

    \textup{4)} System~\eqref{universal_hierarchy_eq}\textup{:}
    \begin{eqnarray*}
      \phi_{0}(A_0) &=& \left(A_0 \, u_t - A_{0,t} \,u, A_0 \, v_t +2 \, A_{0,t}
        \,v \right),
      \\
      \phi_{1}(A_1) &=& \left(A_1, 0\right),
      \\
      \phi_{2}(A_2) &=& \left(0, 2\,A_2 \, u_t + A_{2,t} \,u \right),
      \\
      \phi_{3}(A_3) &=& \left(0, A_3 \right),
      \\
      \phi_{4}(B_0) &=& \left(B_0 \,u_x + B_{0,x} \,y \,u_y, B_0 \,v_x + B_{0,x}
        \,y \,v_y \right),
      \\
      \phi_{5}(B_1) &=& \left(B_1 \,u_y, B_1 \,v_y \right),
      \\
      \phi_{6}(B_2) &=& \left(0, (2 \,B_2 \, u_x- B_{2,x}\, y \,
        u_y)\,u_y^{-3}\right),
      \\
      \phi_{7}(B_3) &=& \left(0, B_3 \,u_y^{-2}\right),
      \\
      \phi_{8} &=& \left(y \, u_y +u, y\, v_y\right),
      \\
      \phi_{9} &=& \left(0, v\right),
      \\
      \phi_{10} &=& \left(0, u_{ttt}\right),
      \\
      \phi_{11} &=& \left(0, (u_y^2 \, u_{ttx}- u_x \,u_y \,u_{tty} + u_x \,
        u_{ty}^2-u_y \, u_{tx} \, u_{ty})\,u_y^{-3} \, \right),
      \\
      \phi_{12} &=& \left(0, (2 \, u_y \, u_{tty}-u_{ty}^2)\,u_y^{-2}\right),
      \\\phi_{13} &=& \left(0, ( 2 \, u_x^2 \, u_y^2 \, (u_x \,
        u_{ttx}-3 \, u_{txy}) -2 \, u_x^4 \, u_y \, u_{tty} -2 \, u_y^3 \,
        (u_{xxy}-2 \, u_x \, u_{txx}) \right.
      \\
      && \qquad \left.  +3 \, u_x \, u_y^2 \, (u_x \,u_{tx}^2 - 2 \, u_{ty}
        \,u_{xx} - 2\,u_{tx} \,u_{xy}) - 4\,u_x^2 \,u_y \,u_{ty} \,(2\,u_x
        \,u_{tx} - 3\,u_{xy}) \right.
      \\
      && \qquad \left.  +u_y^2 \, (3 \, u_{xy}^2 +2 \, u_y \, u_{tx} \,
        u_{xx}) +5 
        \, u_x^4 \, u_{ty}^2 )\,u_y^{-6}\right),
      \\
      \phi_{14} &=& \left(0,
        (2 \, u_y^4 \, u_{xxx}
        +2 \, u_x^4 \, u_y \, (u_y \,u_{ttx}-5 \, u_{tx} \, u_{ty})
        -4 \, u_x^3  \, u_y^2 \, (2\, u_{txy} - u_{tx}^2)
      \right.
      \\
      &&
      \qquad
      \left.
        +6 \, u_x^2 \, u_y^2 \, (u_y u_{txx} - 2 \, u_{ty} \, {u_{xx}}-2 \, u_{tx}
        \, u_{xy})
        -2 \, u_x^5 \, (u_y \, u_{tty} -3 \, u_{ty}^2)
      \right.
      \\
      &&
      \qquad
      \left.
        -6 \, u_y^3 \, (u_x\, u_{xxy}+ u_{xx}  \, u_{xy} - u_x \,  u_{tx} \, u_{xx})
        +4 \, u_x \, u_y \,u_{xy} \,(3 \,u_y \,u_{xy}+5\,u_x^2\,u_{ty})
        )\,u_y^{-7}\right),
      \\
      \phi_{15} &=& \left(0,
        (
        u_x^2 \, (2 \, u_y \, u_{tty}-3 \, u_{ty}^2)
        +u_y^2 \, (2 \, u_{txy}
        -2 \,  u_x \, u_{ttx}
        -u_{tx}^2)
      \right.
      \\
      &&
      \qquad
      \left.
        -2 \, u_y \, u_{ty}\, (u_{xy}-2 \, u_x \,  u_{tx})
        )\,u_y^{-4}\right),
      \\
      \phi_{16} &=& \left(0,
        (
        u_x^3 \, (u_y \, u_{tty}-2 \, u_{ty}^2)
        +u_x \, u_y^2 \, (2 \, u_{txy}-u_{tx}^2)
        -u_y^2 \, (u_x^2 \, u_{ttx}-u_{ty} \, u_{xx}-u_{tx} \, u_{xy})
      \right.
      \\
      &&
      \qquad
      \left.
        -u_y^3 \, u_{txx}
        +3 \,u_x \, u_y \,(u_x \, u_{tx} \, u_{ty}-u_{ty} \, u_{xy})
        )\,u_y^{-5}\right),
    \end{eqnarray*}
    where $A_i=A_i(t)$\textup{,} $B_i=B_i(x)$ are arbitrary functions of their
    arguments\textup{;}

    \textup{5)} System~\eqref{rdDym4D_eq}\textup{:}
    \begin{eqnarray*}
      \phi_0(A_0)  &=& \left(A_0\,u_x - A_{0,x}\,u+A_{0,t}\,z, A_{0} \,v_x +2
        \,A_{0,x}\,v \right), 
      \\
      \phi_1(A_1)  &=& \left(A_1, 0 \right),
      \\
      \phi_2(A_2)  &=& \left(0, 2\,A_2 \, u_x+A_{2,x} \, u-A_{2,t} \, z \right),
      \\
      \phi_3(A_3)  &=& \left(0, A_3 \right),
      \\
      \phi_4(B_0)  &=& \left(B_0 \, u_y, B_0 \, v_y\right),
      \\
      \phi_5(B_1)  &=& \left(0, B_1 \, u_y^{-2}\right),
      \\
      \phi_6  &=& \left(t\, u_t + u, t\,v_t\right),
      \\
      \phi_7  &=& \left(z\, u_z - u, z\,v_z\right),
      \\
      \phi_8  &=& \left(u_z, v_z\right),
      \\
      \phi_9  &=& \left(0, v\right),
      \\
      \phi_{10}  &=& \left(0, u_{xxx}\right),
      \\
      \phi_{11}  &=& \left(0, (2 \, u_y\, u_{xxy} -u_{xy}^2)\,u_y^{-2}\right),
      \\
      \phi_{12}  &=& \left(0, (u_y \, u_{xyy}-u_{xy} \, u_{yy})\,u_y^{-3}\right),
      \\
      \phi_{13}  &=& \left(0, (2 \, t \, u_y^2 \, u_{xxx} -z \,(2 \, u_y\,
        u_{xxy} -u_{xy}^2))\,u_y^{-2}\right),
      \\
      \phi_{14}  &=& \left(0, (t\, u_y \,(2 \, u_y \, u_{xxy}-u_{xy}^2)
        -2 \, z \, (u_y\, u_{xyy}- u_{xy} \, u_{yy}))\,u_y^{-3}\right),
      \\
      \phi_{15}  &=& \left(0, (z^2 \, (u_y\, u_{xyy}-u_{xy} \, u_{yy})
        - t \, z\, u_y\,(2 \, u_y \, u_{xxy} -u_{xy}^2)
        +t^2 \, u_y^3\, u_{xxx}  )\,u_y^{-3}\right),
      \\
      \phi_{16}  &=& \left(0, (2 \, u_y\, u_{yyy} -3 \, u_{yy}^2)\,u_y^{-4}\right),
      \\
      \phi_{17}  &=& \left(0,
        (z\,(2 \, u_y \, u_{yyy}-3 \, u_{yy}^2)
        -2 \, t \, u_y \, (u_y \, u_{xyy}- u_{xy} \, u_{yy}))\,u_y^{-4}\right),
      \\
      \phi_{18}  &=& \left(0,
        (z^2 \,(2 \, u_y \, u_{yyy}-3 \, u_{yy}^2)
        -4 \, t \,z\, u_y \, (u_y \, u_{xyy}- u_{xy} \, u_{yy})
      \right.
      \\
      &&
      \qquad
      \left.
        + t^2  \, u_y^2\,(2 \, u_y \, u_{xxy}-u_{xy}^2)
        )\,u_y^{-4}\right),
      \\
      \phi_{19}  &=& \left(0,
        (
        z^3 \, (2 \, u_y \, u_{yyy}-3 \, u_{yy}^2)
        -6 \, t \, z^2 \, u_y \, (u_y \, u_{xyy}- u_{xy} \, u_{yy})
      \right.
      \\
      &&
      \qquad
      \left.
        + 3 \,t^2 \,z \, u_y^2\,(2 \, u_y \, u_{xxy}-u_{xy}^2)
        -2 \, t^3 \, u_y^4 \, u_{xxx}
        )\,u_y^{-4}\right),
    \end{eqnarray*}
    where $A_i=A_i(t,x)$\textup{,}  $B_i=B_i(y,z)$ are
    arbitrary functions of their arguments.

    All the functions above are assumed to be smooth and the second subscript
    \textup{(}as in\textup{,} say\textup{,}~$A_{3,t}$\textup{)} denotes the
    corresponding derivative.
\end{proposition}

Thus, we have the following higher symmetries:
\begin{align*}
  &\text{For Systems~\eqref{Ovsienko_eq}, \eqref{Veronese_web_eq},
    \eqref{universal_hierarchy_eq}:\qquad
  }&\phi_{10},\dots,\phi_{16},\\
  &\text{For System~\eqref{Pavlov_eq}:}&\phi_{11},\dots,\phi_{17},\\
  &\text{For System~\eqref{rdDym4D_eq}:\qquad }&\phi_{10},\dots,\phi_{19}.\\
\end{align*}

\section{Bivectors on the deformed cotangent covering}
\label{sec:bivect-deform-cotang}
Let~$\ES{E}$ be an equation and~$\tau^*\colon\ES{T}^*\ES{E}\to\ES{E}$ be its
tangent covering. As it was mentioned in
Section~\ref{sec:basics},~$\ES{T}^*\ES{E}$ is an Euler-Lagrange
equation. Let~$\ES{T}^*\ES{E}$ be given by~$F(x,u)=0$,
$\ell_{\ES{E}}^*(x,u,v)=0$. Then we say that the system
\begin{equation}\label{eq:1}
  \tilde{\ES{E}}\colon\qquad \ell_{\ES{E}}^*(x,u,v)+G(x,u,v)=0,\quad
  F(x,u)+H(x,u,v)=0
\end{equation}
is a \emph{Lagrangian deformation} (\ld) if~$\tilde{\ES{E}}$ is also an
Euler-Lagrange equation. We know at least two meaningful examples of this
construction: (i) The system
\begin{equation*}
   u_{ty} = u_{xy} u_x - u_{xx} u_y,\qquad
    v_{ty} = 2 (u_{xx} v_y - u_{xy} v_x) + u_x v_{xy} - u_y v_{xx} -2 (u_{xx}
    u_y + 2 u_{xy} u_x),
\end{equation*}
see~\cite{Ovsienko2010}, is an \ld\ of~\eqref{Ovsienko_eq}; (ii) The Dunajski
equation,~\cite{Dunajski2002},
\begin{equation*}
  \Theta_{wx} + \Theta_{zy} + \Theta_{xx}\Theta_{yy} - \Theta_{xy}^2 = f,\qquad
  f_{xw} + f_{yz} + \Theta_{yy}f_{xx} + \Theta_{xx}f_{yy} -
  2\Theta_{xy}f_{xy} = 0
\end{equation*}
is an \ld\ of the $2$nd Heavenly Equation.

The identical map from~$\cosym(\tilde{\ES{E}})$ to~$\sym(\tilde{\ES{E}})$ is
always a bivector on any \ld~$\tilde{\ES{E}}$ and we pose the following
problem: \emph{Given an equation~$\ES{E}$\textup{,} can~$\ES{T}^*\ES{E}$ be
  deformed in such a way that the resulting \ld~$\tilde{\ES{E}}$ will admit
  nontrivial bivectors}?

We studied \ld s for Equations~\eqref{Ovsienko_eq},
\eqref{universal_hierarchy_eq}, and~\eqref{rdDym4D_eq} and obtained the
following result:
\begin{proposition}\label{prop:bivect-deform-cotang-1}
  Equations~\eqref{Ovsienko_eq}\textup{,}
  \eqref{universal_hierarchy_eq}\textup{,} and~\eqref{rdDym4D_eq} possess
  two-parameter families of \ld s with~$H=0$ and~$G=\delta
  L_{\kappa_1,\kappa_2}/\delta u$ and three-dimensional spaces of
  bivectors~$\Delta=\alpha_0\id+\alpha_1\Delta^1+\alpha_2\Delta^2$\textup{,}
  $\Delta^i=(\Delta_{jk}^i)$\textup{,} $i,j,k=1,2$\textup{,} where
  \begin{align*}
    \intertext{For Equation~\eqref{Ovsienko_eq}\textup{:}}
    L_{\kappa_1,\kappa_2}&=(\kappa_1u_y+\kappa_2)u_x^2,\\[2pt]
    \Delta_{11}^1&=u_yD_x^{-1}\circ u_y^{-2}\circ D_y,\\
    \Delta_{12}^1&=0,\\
    \Delta_{21}^1&=(v_y-\kappa_2)D_x^{-1}\circ u_y^{-2}\circ D_y +
    u_y^{-2}D_x^{-1}\circ(v_y-\kappa_2)\circ D_y,\\
    \Delta_{22}^1&=u_y^{-2}D_x^{-1}\circ
    u_y\circ D_y ,\\[2pt]
    \Delta_{11}^2&=\frac{1}{2}u_x + \frac{1}{2}D_x^{-1}\circ(D_t - 2u_xD_x),\\
    \Delta_{12}^2&=0,\\
    \Delta_{21}^2&=v+\kappa_1uD_x -
    \frac{1}{2}D_x^{-1}\circ(2\kappa_1uD_x+v_{xx}) -
    \kappa_1D_x^{-1}\circ (D_t-2u_xD_x),\\
    \Delta_{22}^2&=\frac{1}{2}uD_x-u_x +\frac{1}{2}D_x^{-1}\circ (D_t-uD_x^2).
    \intertext{For Equation~\eqref{universal_hierarchy_eq}\textup{:}}
    L_{\kappa_1,\kappa_2}&=u_t(\kappa_1u_y+\kappa_2u_x),\\[2pt]
    \Delta_{11}^1&=uD_t^{-1}\circ u_y^{-2}\circ D_y,\\
    \Delta_{12}^1&=0,\\
    \Delta_{21}^1&=(v_y+\kappa_2)D_t^{-1}\circ u_y^{-2}\circ D_y +
    u_y^{-2}D_t^{-1}\circ
    (v_y+\kappa_2) \circ D_y,\\
    \Delta_{22}^1&=u_y^{-2}D_t^{-1}\circ u_y\circ D_y,\\[2pt]
    \Delta_{11}^2&=uD_t-D_y^{-1}\circ (u_{yt} +uD_yD_t),\\
    \Delta_{12}^2&=0,\\
    \Delta_{21}^2&=-2vD_t - D_y^{-1}\circ (v_{yt} + 2\kappa_2 D_t - 2vD_yD_t),\\
    \Delta_{22}^2&=uD_t + D_y^{-1}\circ (uD_yD_t-2u_{yt}).  \intertext{For
      Equation~\eqref{rdDym4D_eq}\textup{:}}
    L_{\kappa_1,\kappa_2}&=u_x(\kappa_1u_z+\kappa_2u_y),\\[2pt]
    \Delta_{11}^1&=-uD_x + D_z^{-1}\circ(u_{xz}+D_t+uD_xD_z),\\
    \Delta_{12}^1&=0,\\
    \Delta_{21}^1&=-D_z^{-1}\circ(v_{xz} + 2(v_z+\kappa_2)D_x),\\
    \Delta_{22}^1&=-uD_x + 2u_x +
    D_z^{-1}\circ (u_{xz}+D_t+uD_xD_z),\\[2pt]
    \Delta_{11}^2&=u_yD_x^{-1}\circ u_y^{-2}\circ D_y,\\
    \Delta_{12}^2&=0,\\
    \Delta_{21}^2&=-u_y^{-2}D_x^{-1}\circ(v_y+\kappa_1)\circ D_y -
    (v_y+\kappa_1)D_x^{-1}\circ u_y^{-2}\circ D_y,\\
    \Delta_{22}^2&=u_y^{-2}D_x^{-1}\circ u_yD_y.
  \end{align*}
  \end{proposition}

\section{Concluding remarks}
\label{sec:concluding-remarks}

\begin{remark}
  The identical bivector in Equation~\eqref{Ovsienko_eq}, after transforming
  the system to the evolutionary form, corresponds to the first Hamiltonian
  operator from~\cite{Ovsienko2010}, while the bivector~$\Delta^1$ corresponds
  to the second one.
\end{remark}

\begin{remark}
  It can be shown that the \ld s for Equations~\eqref{universal_hierarchy_eq}
  and~\eqref{rdDym4D_eq} presented above are in a sense trivial, because the
  deformed equations are transformed to the initial cotangent space by a point
  transformation. As for Equation~\eqref{Ovsienko_eq}, there exists a
  covering~$\tau\colon\tilde{\tilde{\ES{E}}}\to\tilde{\ES{E}}$ (that
  corresponds to a differential substitution) such
  that~$\tilde{\tilde{\ES{E}}}$ is equivalent to~$\ES{T}^*\ES{E}$.
\end{remark}

\begin{remark}
  As it was mentioned above, all computations were done using the
  \textsc{Jets} (\textsc{Maple~15}),~\cite{Jet,Marvan2009}, software.  We used
  the IBM Blade cluster consisting of 3 working machines, each of them
  quad-core Intel Xeon: one E5460, 2.66 GHz, 56 GB RAM two E5430, 2.66 GHz, 16
  GB RAM.  The classification tree contained~$3446$ branches, $15$ of them
  were abandoned due to lack of memory. The results presented are due to
  successful processing of $11$ branches only. Other branches are to be
  processed in future. We also plan to consider \ld s of a
  more general nature.
\end{remark}

\begin{remark}
  Actually, presenting results of Proposition~\ref{prop:bivect-deform-cotang-1}
  in the operator form is not complete and is due to the common tradition
  only. It is more adequate to deal with them as with B\"{a}cklund
  transformations between the tangent (see~\cite{KrasilshchikVerbovetsky2011})
  and cotangent spaces of the equation at hand (cf.~\cite{Morozov2013}), but
  the operator form seems to be ``more visual''.
\end{remark}

\ack{The work of H.~Baran and P.~Voj{\v{c}}{\'{a}}k was supported by RVO
  funding for I\v{C}47813059. The work of H.~Baran was also supported by Czech
  Science Foundation (GA\v{C}R) project
  P201/11/0356. I.~Krasil${}^\prime$shchik was supported by the European
  Social Fund under the project CZ.1.07/2.3.00/20.0002.}


\section*{References}

\begin{thebibliography}{99}

\bibitem{Jet} Baran H, Marvan M  Jets. A software for differential calculus on
  jet spaces and diffieties. {\tt http://jets.math.slu.cz}

\bibitem{Blaszak} B{\l}aszak M 2002 Classical R-matrices on Poisson algebras
  and related dispersionless systems \emph{Phys.\ Lett.} A \textbf{297} 191--5

\bibitem{BurovskyFerapontovTsarev2010} Burovskiy P A, Ferapontov E V, Tsarev S
  P 2010 Second order quasilinear PDEs and conformal structures on in
  projective space \emph{Int.\ J.\ Math.}  \textbf{21} 799--841

\bibitem{Dunajski2002} Dunajski M 2002 Anti-self-dual four-manifolds with a
  parallel real spinor \emph{Proc.\ Roy.\ Soc.\ Lond.}\ \textbf{A 458}
  1205--1222

\bibitem{Dunajski2004} Dunajski M 2004 A class of Einstein-Weil spaces
  associated to an integrable system of hydrodynamic type \emph{J.\ Geom.\
    Phys.}  \textbf{51} 126--37

\bibitem{Marvan2009} Marvan M 2009 Sufficient set of integrability conditions
  of an orthonomic system \emph{Foundations of Comp. Math.} \textbf{9} 651--74

\bibitem{MarvanSergyeyev2012} Marvan M, Sergyeyev A 2012 Recursion operators
  for dispersionless integrable systems in any di\-men\-si\-ons \emph{Inverse
    Problems} \textbf{28} 025011

\bibitem{KrasilshchikVerbovetsky2011} Krasil${}^{\prime}$shchik J, Verbovetsky
  A 2011 Geometry of jet spaces and integrable systems \emph{J.\ Geom.\ Phys.}
  \textbf{61} 1633--74

\bibitem{KrasilshchikVerbovetskyVitolo2012} Krasil${}^{\prime}$shchik I S,
  Verbovetsky A M, Vitolo R 2012 A unified approach to computation of
  integrable structures \emph{Acta Appl.\ Math.} \textbf{120} 199--218

\bibitem{AlonsoShabat2004} Mart{\'{\i}}nez Alonso L, Shabat A B 2004
  Hydrodynamic reductions and solutions of a universal hierarchy \emph{
    Theor.\ Math.\ Phys.} \textbf{140} 1073--85

\bibitem{Mikhailov} Mikhailov A V (Ed) 2009 \emph{Integrability} Lecture Notes
  in Physics \textbf{767} (Berlin: Springer)

\bibitem{Morozov2009} Morozov O I 2009 Contact integrable extensions of
  symmetry pseudo-groups and coverings of (2+1) dispersionless integrable
  equations \emph{J.\ Geom.\ Phys.} \textbf{59} 1461--75

\bibitem{Morozov2012} Morozov O I 2013 A Recursion Operator for the Universal
  Hierarchy Equation via Cartan's Method of Equivalence, Central Euro. J. Math., accepted, arXiv:1205.5748
  [nlin.SI]

\bibitem{Morozov2013} Morozov O I 2013 A Four-Dimensional Generalization of
  the Integrable rdDym Equation \emph{Preprint} arXiv:1309.4993 [nlin.SI]

\bibitem{Ovsienko2010} Ovsienko V 2010 Bi-Hamiltonian nature of the equation
  $u_{tx} = u_{xy}\,u_{y} - u_{yy}\, u_{x}$ \emph{Adv.\ Pure Appl.\ Math.} \textbf{
    1} 7--17

\bibitem{OvsienkoRoger2008} Ovsienko V, Roger C 2007 Looped cotangent Virasoro
  algebra and non-linear integrable systems in dimension $2+1$ \emph{
    Comm.\ Math.\ Phys.} \textbf{273} 357--88

\bibitem{Pavlov2003} Pavlov M V 2003 Integrable hydrodynamic chains \emph{
    J.\ Math.\ Phys.} \textbf{44} 4134--56

\bibitem{Pavlov2006} Pavlov M V 2006 The Kupershmidt hydrodynamics chains and
  lattices 2006 \emph{Intern.\ Math. Research Notes} \textbf{2006} 46987

\bibitem{Vinogradov:SSLFCLLTNT}  Vinogradov A M 1984 The
    $\ES{C}$-spectral sequence, {L}agrangian formalism, and conservation
    laws. \textup{I}.\ {T}he linear theory.  \textup{II}.\ {T}he nonlinear
    theory \emph{J.\ Math.\ Anal.\ Appl.} \textbf{100} 1--129.

\bibitem{Vinogradov1989} Vinogradov A M 1989 Symmetries and conservation laws
  of partial differential equations: basic notions and results \emph{Acta
    Appl.\ Math.} \textbf{15} 3--21

\bibitem{Zakharevich} Zakharevich I 2000 Nonlinear wave equation, nonlinear
  Riemann problem, and the twistor transform of Veronese webs \emph{Preprint}
  math-ph/0006001

\bibitem{Zakharov} Zakharov V E (Ed) 1992 \emph{What is integrability?}
  Springer series in Nonlinear Dynamics (Berlin: Springer)
\end{thebibliography}
\end{document}